\documentclass[prl, letterpaper, showpacs,twocolumn, floatfix, superscriptaddress,
balancelastpage, 10pt]{revtex4} \usepackage{graphicx, dcolumn,color,booktabs}
\usepackage{microtype} \usepackage{amssymb, amsmath}
\usepackage[colorlinks,plainpages=false,linkcolor=black,urlcolor=blue,citecolor=blue,pdfpagemode=UseNone,pdfstartview=FitH]{hyperref}

\newcommand
\sample{Sr$_{\scriptsize{\textup{2}}}$RuO$_{\scriptsize{\textup{4}}}${}}

\renewcommand{\figurename}{Fig.} \renewcommand{\tablename}{Table}
\makeatletter\renewcommand{\fnum@figure}[1]{\figurename~\thefigure}\makeatother
\makeatletter\renewcommand{\fnum@table}[1]{\tablename~\thetable.}\makeatother

\begin{document} \title{Effective tight-binding model for renormalized band
structure of {\sample}}

\author{V.\,B.\,Zabolotnyy}
\author{D.\,V.\,Evtushinsky} \affiliation{Institute for
Solid State Research, IFW-Dresden, P. O. Box 270116, D-01171 Dresden, Germany}
\author{A.\,A.\,Kordyuk} \affiliation{Institute of Metal Physics of National
Academy of Sciences of Ukraine, 03142 Kyiv, Ukraine} \affiliation{Institute for
Solid State Research, IFW-Dresden, P. O. Box 270116, D-01171 Dresden, Germany}
\author{T.\,K.\,Kim\footnote{Present address: Diamond Light Source Ltd., Didcot,
Oxfordshire, OX11 0DE, United Kingdom}} \affiliation{Institute for Solid State
Research, IFW-Dresden, P. O. Box 270116, D-01171 Dresden, Germany}
\author{E.\,Carleschi} \affiliation{Department of Physics, University of
Johannesburg, P. O. Box 524, Auckland Park 2006, South Africa}
\author{B.\,P.\,Doyle} \affiliation{Department of Physics, University of
Johannesburg, P. O. Box 524, Auckland Park 2006, South Africa}
\author{R.\,Fittipaldi} \affiliation{CNR-SPIN, and Dipartimento di Fisica ``E. R.
Caianiello'', Universit\`{a} di Salerno, I-84084 Fisciano (Salerno) Italy}
\author{M.\,Cuoco} \affiliation{CNR-SPIN, and Dipartimento di Fisica ``E. R.
Caianiello'', Universit\`{a} di Salerno, I-84084 Fisciano (Salerno) Italy}
\author{A.\,Vecchione} \affiliation{CNR-SPIN, and Dipartimento di Fisica ``E. R.
Caianiello'', Universit\`{a} di Salerno, I-84084 Fisciano (Salerno) Italy}
\author{S.\,V.\,Borisenko} \affiliation{Institute for Solid State Research,
IFW-Dresden, P. O. Box 270116, D-01171 Dresden, Germany} \date{\today}

\begin{abstract} We derive an effective quasiparticle tight-binding model which is
able to describe with high accuracy the low-energy electronic structure of
Sr$_2$RuO$_4$ obtained by means of low temperature angle resolved photoemission
spectroscopy. Such approach is applied to determine the momentum and orbital
dependent effective masses and velocities of the electron quasiparticles close to
the Fermi level. We demonstrate that the model can provide, among the various
computable physical quantities, a very good agreement with the specific heat
coefficient and the plasma frequency. Its use is underlined as a realistic input in
the analysis of the possible electronic mechanisms related to the superconducting
state of Sr$_2$RuO$_4$.

\end{abstract} 

\preprint{\textit{xxx}} \maketitle

Since its discovery, the nature of the superconducting state of {\sample} remains
in the focus of the solid state research\,\cite{Carlo323, Puetter27010, Maeno011009, Wysokinski077004}. An accurate description of the low energy electronic structure is a fundamental step for understanding the collective properties of complex
materials. This is also the case for the superconducting phase of Sr$_2$RuO$_4$.
There are generally two ways to get access at the electronic structure of a given
material. On one side, \emph{ab initio} density functional theory (DFT) can provide quasiparticle spectrum at all energies, although it is known to be not suitable for properly accounting the effects of electron correlations. To this end, DFT
calculations are often taken as a platform for a more elaborate treatment of
correlation effects as, for instance, in DFT+DMFT (dynamical mean-field theory)
approaches, or other many-body theories. Such methods, in the attempt to build up
an accurate quantitative description of correlated materials, usually includes the
Coulomb interaction within tight-binding (TB) models based on a localized Wannier
basis from the DFT states. The complexity in dealing with the high and low energy
sector of correlated materials on equal footing leads to deviations between the
theoretical predictions and the experimental observations. These can manifest themselves, for instance, in the difficulty to capture the observed band renormalization, to quantitatively reproduce the relative band positions\,\cite{Geck046403}, \emph{etc}.

On the other hand, there are different experimental methods to probe directly and
indirectly the electronic structure. For instance, the thermodynamical properties
can provide average information on the physical quantities at the Fermi level (FL) such as density of states. Otherwise, by means of de Haas--van Alphen or Shubnikov--de
Haas measurements via the analysis of the resonance frequencies of the cyclotron
motion it is possible to map the Fermi surface and to extract the effective masses
at the FL, assuming that suitable conditions for the applied magnetic field and the degree of purity of the samples are given. For the Compton scattering probe, which recently gained popularity with layered superconductors, one has to
face the reconstruction of a 2D electron density from a set of experimentally
measured Compton profiles\,\cite{Hiraoka100501, Hiraoka094511, Mijnarends2381,
AlSawai115109, Utfeld064509, Sakurai06052011}. In this framework, in terms of band
mapping\,\cite{Krasovskii045432}, angle-resolved photoelectron spectroscopy (ARPES) appears to be the most direct momentum and energy resolving technique for determination of the
electronic structure.

Concerning the Sr$_2$RuO$_4$, though for the first ARPES measurements it was not
easy to disentangle the contributions of the surface states from the bulk
ones\,\cite{Yokoya3009}, the improvement of the experimental analysis allowed to get a general agreement between photoemission and bulk probes\,\cite{Bergemann639,
Hiraoka100501}. Interestingly, the recent observation of an anomalous splitting of
the $\beta$ surface bands renewed interest in the study of Sr$_2$RuO$_4$ electronic structure\,\cite{Zabolotnyy063039}. While various reports on integrated quantities
(like average Fermi velocities or effective masses) characterizing the band
structure of {\sample} are available in the literature, a detailed quantitative
description of the low energy electronic structure of {\sample} as measured by
ARPES is still missing. In this paper, starting from low temperature high
resolution ARPES observations, we aim at providing an effective TB model
to quantitatively describe the dispersion of the renormalized low energy
quasiparticles of {\sample},
following an approach that is similar to what has  already been done for the
layered dichalcogenides $2H$-TaSe$_2$ and $2H$-NbSe$_2$\,\cite{Inosov125112,
Inosov125027}.

Effective TB models, i.e. a representation of the electronic structure
within a certain energy region close to the FL in terms of atomic-like
orbitals, is a powerful method often used to analyze the essential mechanisms
governing the physical behavior of complex materials. Moreover, one of the basic
advantages of a TB model is that it allows the band structure to be computed
on very fine meshes in the Brillouin zone at low computational cost, which,
furthermore, greatly facilitates calculation of transport, superconducting and other properties determined by peculiarities of the Fermi surface and the dispersion of low energy electronic bands\,\cite{Mazin5223}.

TB models with the corresponding sets of parameters as derived from the
first-principles calculations of \sample{} have been reported
earlier\,\cite{Mazin733, Liebsch1591, Morr5978, Mishonov305, Mazin5223, Noce2659,
Noce19971713}, and used to calculate the magnetic response\,\cite{Braden064522,
Morr5978}, the Hall coefficient\,\cite{Noce2659} and the photoemission spectra\,\cite{Liebsch1591}. Unlike the previous examples, where unrenormalized band
structure was captured,  TB models were also successfully applied to parameterize
the dynamics of quasiparticles, as in the case of graphene\,\cite{Grueneis205425},
for the reconstructed diamond surface C(111)2$\times$1\,\cite{Marsili205414} or in
iron arsenides\,\cite{Beaird140507}. Here we combine our experimental data with a
quasiparticle tight-binding approach to produce an accurate description of
quasiparticle dispersion in  single layer ruthenate Sr$_2$RuO$_4$ in the vicinity of the FL.

High-quality {\sample} single crystals used in this work have been grown by the
flux-feeding floating-zone technique with Ru self-flux\,\cite{Fittipaldi70180,
Mao20001813}. The composition and structure of the samples have been characterized
by X-ray  and electron backscatter diffraction.  All the diffraction peaks had the
expected (001) Bragg reflections of the {\sample} phase, confirming the absence of
any spurious phase. The purity of the crystals is supported by a.c. susceptibility
and resistivity measurements demonstrating a narrow superconducting transition with
$T_\textup{c}$=1.34\,K, which is a signature of a low impurity
concentration\,\cite{Kikugawa237}. Photoemission data were collected at the BESSY
1$^3$ ARPES station equipped with a SCIENTA R4000 analyzer and a Janis $^3$He
cryostat\,\cite{Borisenko720159, BorisenkoJove}. Further details on the experimental geometry can be found elsewhere\,\cite{Zabolotnyy024502, Inosov212504}.

\begin{figure}[t] \includegraphics[width=0.9\columnwidth]{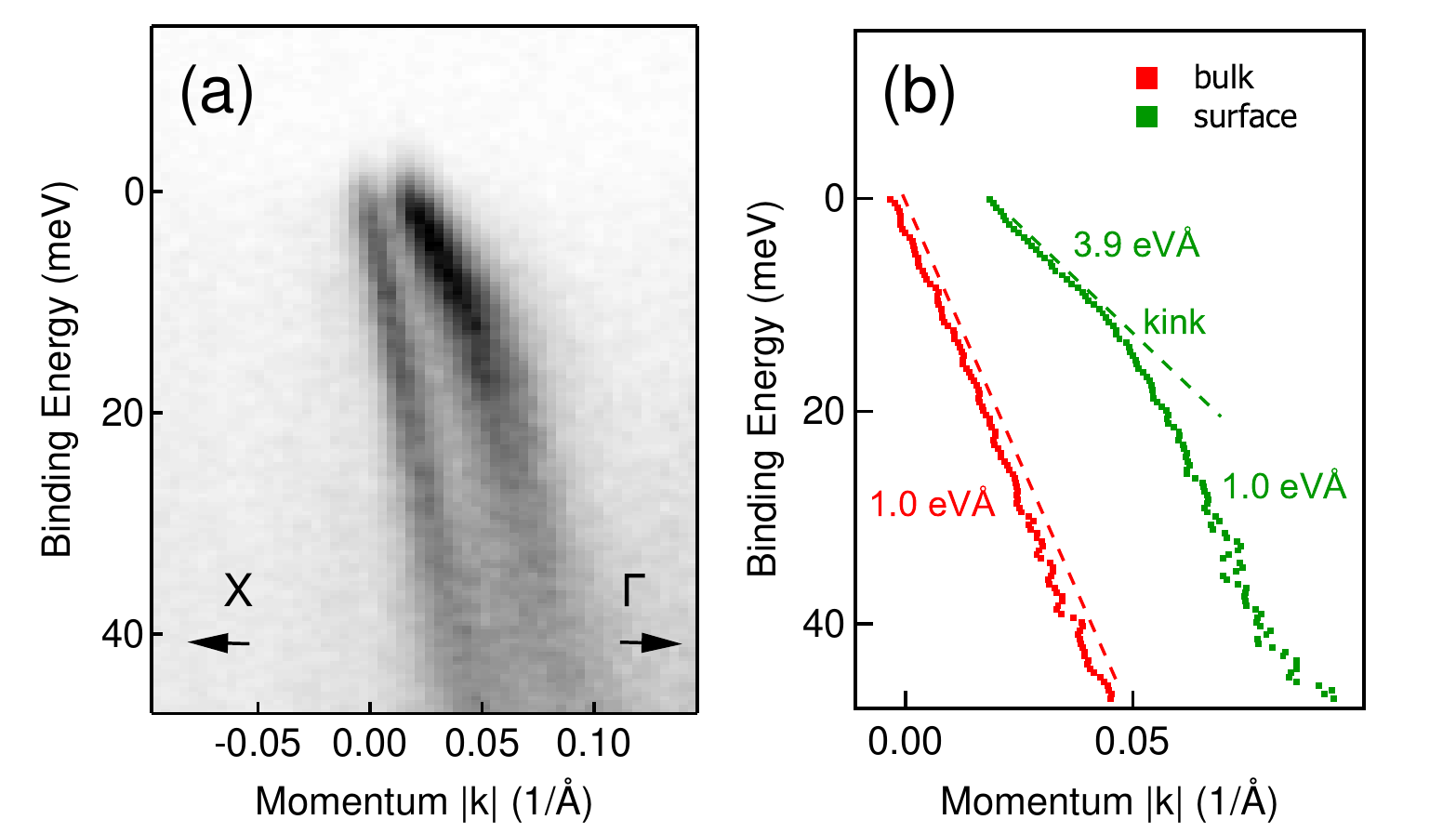} \caption{ Surface
and bulk $\alpha$ bands (a) and their MDC dispersions (b). The projections of the
band velocities were estimated by fitting a line to a straight segments of
experimental dispersion. Since the cut is not perpendicular to the Fermi surface
locus, the total in-plane velocities are actually larger. } \label{alphacut}
\end{figure}

Before presenting the modeling of the {\sample} electronic structure it is worth
pointing out a few aspects which have to be considered with care in the attempt of
deriving a TB description of the experimental data\,\cite{Inosov125112,
Inosov125027, Kordyuk064504, Evtushinsky147201}. Indeed, electronic structure of
{\sample} as seen in photoemission experiment can be regarded as a superposition of
two sets of features, one corresponding to the bulk bands, and the other one to the surface bands\,\cite{Ingle205114, Zabolotnyy063039}. While the momentum disparity
between the corresponding surface and bulk features is comparatively small at the
FL, the difference becomes notable at higher binding energies because of
the unequal renormalization of the surface and bulk bands\,\cite{Ingle205114,
Zabolotnyy024502, Zabolotnyy063039}. To illustrate this issue in
Fig.\,\ref{alphacut} we show a cut through the $\alpha$ pocket, where the surface
and bulk $\alpha$ bands are well resolved, so that their MDC dispersions can be fit
and traced down to about 50\,meV in binding energy. We find that the velocity of
the bulk band projected on the cut direction is about 1\,eV$\cdot$\AA, and does not
vary much within the first 50\,meV below the FL. However, for the surface band,
contrary to the expectations expressed in Ref.\,\onlinecite{Ingle205114}, we find
an abrupt change in the band velocity located at  about 17\,meV binding energy.

Such a kink in the band dispersion typically signals the occurrence of a coupling
between electrons and bosonic modes, which at these energies are typically ascribed
to phonons\,\cite{Engelsberg993,Sandvik094523, Kordyuk134513, Rahn224532}.
Considering the evidence for a strong electron phonon coupling in {\sample}, which is based on the neutron data by Braden \emph{et al.}\,\cite{Braden1236, Braden014505} and theoretical calculations\,\cite{Bauer395701, Wang172503}, it is interesting to have a closer look at this issue. One may notice that up to 4\,THz ($\sim$16.5\,meV) there are only acoustic phonon branches, and in the range 4--5\,THz  weakly dispersing optical phonons of various symmetries are present, which seem to be a good candidate to cause the observed kink in the band dispersion. Such a variation in the electron--phonon coupling for the bulk and surface bands may seem surprising at first.  However, the lower symmetry of the local ionic environment is likely to account for the enhanced electron--phonon coupling at the metal surface\,\cite{Plummer2003251, Mazin198893}. This dichotomy also helps to clarify the difference between the Ingle \textit{et al.}\,\cite{Ingle205114} and Iwasawa
\textit{et al.}\,\cite{Iwasawa104514} reports, who showed a practically flat
dispersion for the $\alpha$ band, on one side, and the kink reported by Aiura
\textit{et al.}\,\cite{Aiura117005} and Kim \textit{et al.}\,\cite{Kim2011556} on the other side. In view of the current data we believe the latter two experiments must have been performed under conditions of a dominating surface component.

\begin{figure*}[t]
\includegraphics[width=0.85\textwidth]{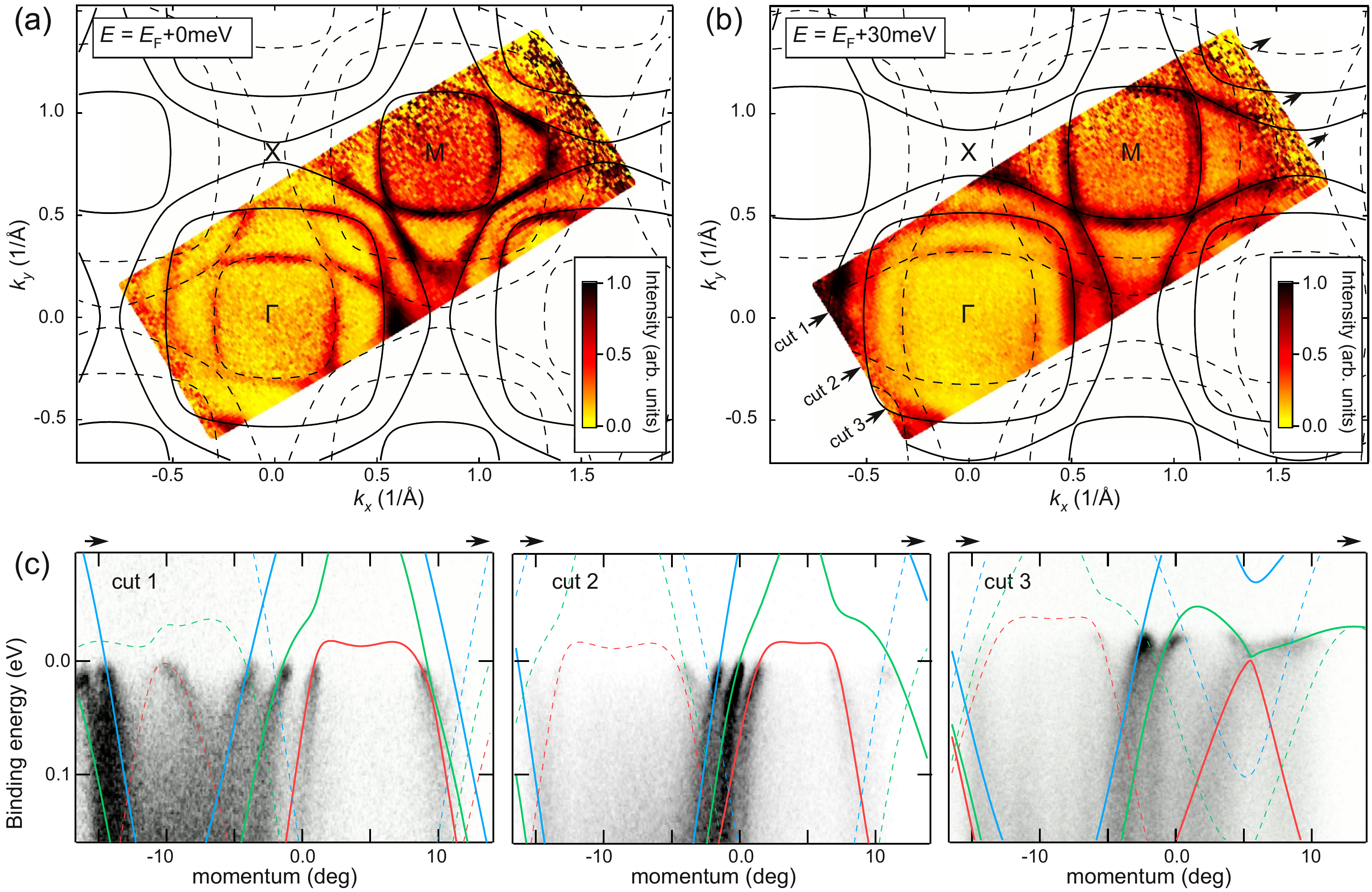}\\
\caption{(a) Experimental Fermi surface of Sr$_2$RuO$_4$ with superimposed TB contours. (b) Intensity distribution similar the Fermi surface map shown in (a) but taken 30\,meV below the FL. (c) Comparison of experimental intensity distribution for several energy--momentum cuts with fitting quasiparticle dispersion. Cuts position in momentum space is marked by small arrows in panel (b).}
\label{Ruth1}
\end{figure*}

There are two outcomes from this observation. The first one, mainly pertaining to
the current study, is that  when  constructing  any TB fit intended to
describe the bulk band structure of {\sample}, one always has to pick the band with a higher Fermi velocity from the two close bulk and surface features. A higher
Fermi velocity for the bulk counterparts as compared to surface ones has also been
observed in other layered superconductor YBaCu$_2$O$_{7-\delta}$\,\cite{Zabolotnyy064507}, which brings about the second
outcome, apparently affecting band renormalization studies performed with ARPES in
general. The point is that the momentum splitting between the surface and bulk
bands might be negligibly small, hence treating an unresolved  composite feature as a single one may lead to an underestimated Fermi velocity (overestimated
renormalization) as contrasted to the true bulk values.

Let us consider the effective TB model. In comparison to many other
layered superconductors {\sample} is known to have a relatively weak $k_z$
dispersion\,\cite{Markiewicz054519, Eschrig104503, Takeuchi227004, Bansil012503,
Rossnagel073102}, which is still further reduced by the spin--orbit
interaction\,\cite{haverkort026406}. Therefore, in choosing an appropriate
TB model for {\sample} we neglect the $k_z$ dispersion and follow the
basic formulation already proposed by Ng et al. as well as by other
authors\,\cite{Ng473, Puetter081105, Puetter27010, Lee184403}. In this framework, the TB Hamiltonian can be expressed as follows:
\begin{equation}
 H = \sum\limits_{\vec{k}, s}\psi_s^\dagger(\vec{k}) \hat{A}(\vec{k})
 \psi_s(\vec{k}) + \rm{h.c.}{,}
\end{equation}
where $\psi_s(\vec{k}) = [d^{yz}_s(\vec{k}),d^{xz}_s(\vec{k}),
d^{xy}_{-s}(\vec{k}) ]^T$ indicates the basis with a three-component spinor and the
matrix $\hat{A}(\vec{k}, s)$ is given by
\begin{equation} \hat{A}(\vec{k}) =
\left(
\begin{array}{ccc} \epsilon^{yz}_{\vec{k}} - \tilde{\mu} &\,
\epsilon^{\mathrm{off}}_{\vec{k}} + i\lambda &\,\,  - \lambda \\
\epsilon^{\mathrm{off}}_{\vec{k}} - i\lambda &\, \epsilon^{xz}_{\vec{k}} -
\tilde{\mu}&\,\,\, i \lambda \\ -\lambda &\, -i\lambda   &\,
\epsilon^{xy}_{\vec{k}} - \tilde{\mu}
\end{array}
\right),   \mathrm{and}
\end{equation} {\noindent
$\epsilon^{yz}_{\vec{k}} = - 2 \tilde{t}_2   \cos(k_x)   -
2 \tilde{t}_1    \cos(k_y);$\\ $\epsilon^{xz}_{\vec{k}}   =   -   2  \tilde{t}_1
\cos(k_x)  -  2 \tilde{t}_2   \cos(k_y);$\\ $\epsilon^{xy}_{\vec{k}} = - 2
\tilde{t}_3 (\cos(k_x) + \cos(k_y))  - 4 \tilde{t}_4 \cos(k_x) \cos(k_y) -2
\tilde{t}_5(\cos( 2 k_x) + \cos(2 k_y));\\
\epsilon^{\mathrm{off}}_{\vec{k}} = - 4\tilde{t}_6 \sin(k_x) \sin(k_y)$.}

\vspace{2mm}

In Fig.\,\ref{Ruth1} we fit the model parameters in order to optimally reproduce
the experimental data in an energy window close to the FL. Panel (a) shows
the TB-model Fermi surface contours superimposed over the experimental data. The
effective electronic parameters which provide the best description for the
dispersion of the low energy quasiparticles in Sr$_2$RuO$_4$ can be summarized in
the following table:
\begin{table}[h]
\begin{tabular}{cccccccc} \firsthline
$\phantom{\widehat{|}}\tilde{\lambda}$ & $\tilde{t}_1$ & $\tilde{t}_2$ &
$\tilde{t}_3$ & $\tilde{t}_4$ & $\tilde{t}_5$ & $\tilde{t}_6$ & $\tilde{\mu}$\\
\hline
 0.032 &  0.145 &  0.016 &  0.081 &  0.039 &  0.005 &  0.000 & 0.122\\
\lasthline
\end{tabular}
\end{table} 
\begin{figure}[t]
\center
\includegraphics[width=0.99\columnwidth]{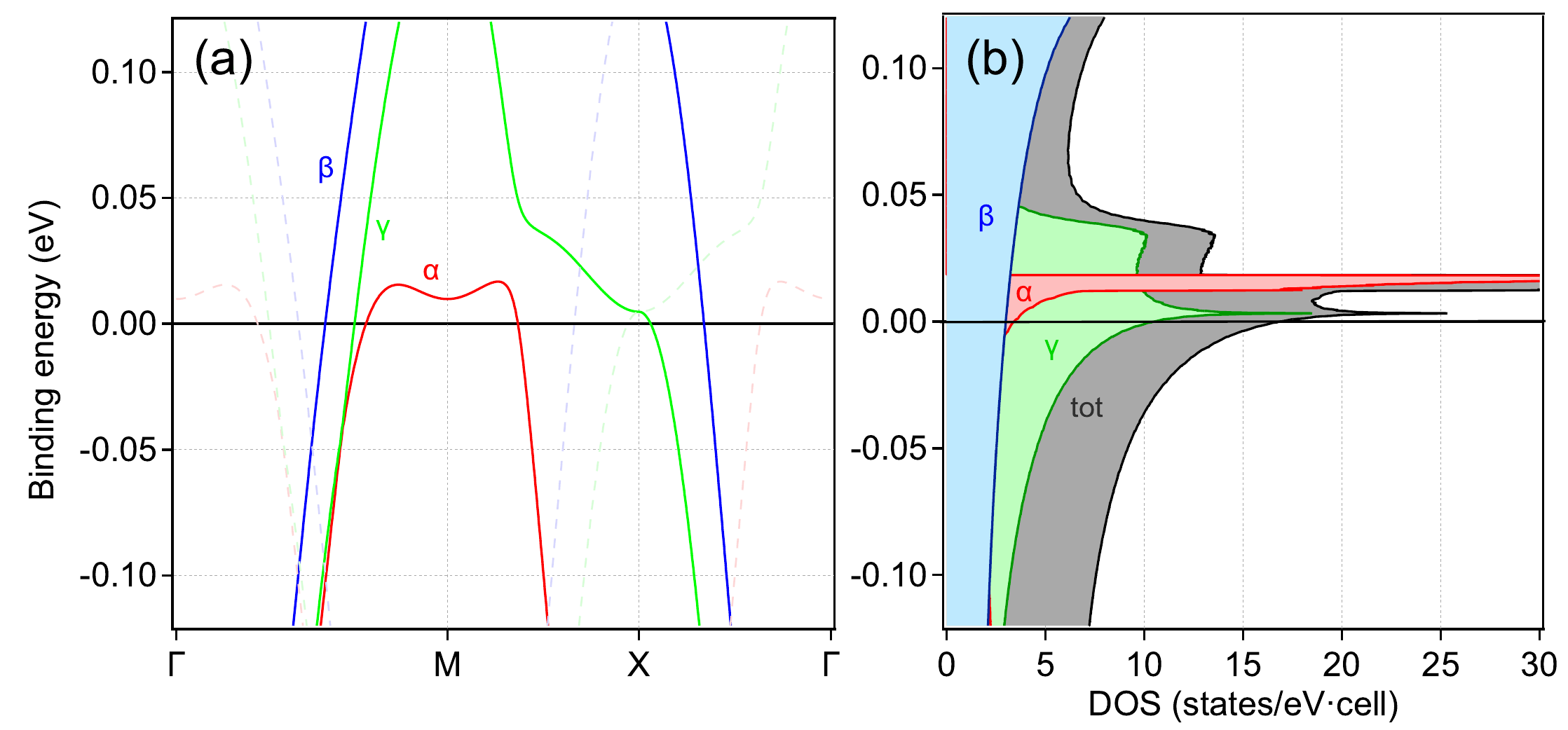}\\
\caption{Dispersion of the quasiparticle TB bands (a) and the derived quasiparticle density of states (b) in the vicinity of the FL.}
\label{Ruth2}
\end{figure} %

\vspace{-0.2cm}
{Having at hand simple equations describing the band dispersion} it is easy to calculate the density of states (DOS) and make an estimate for the electron count and electronic specific heat\,\cite{Stockert224512}. To calculate DOS we used tetrahedron method of Lehmann and Taut\,\cite{Lehmann540211}. The results are shown in Fig.\,\ref{Ruth2}.
The estimated density of states at the FL for the three low energy bands are as follows: $g_\alpha\approx3.4$,    $g_\beta\approx3.0$ and    $g_\gamma\approx10.5$\,states/eV$\cdot$molRu, which totals to    $g\approx16.9$\,states/eV$\cdot$molRu and translates to the Sommerfeld coefficient    of 40\,mJ/K$^2\cdot$molRu. The obtained number is in agreement  with the    experimentally measured values ranging from 29 to 45.6 mJ/K$^2\cdot$molRu\,\cite{Carter17184, Nishizaki560, Nishizaki13401, Nishizaki572, Yoshiteru1405,    Maeno532, Neumeier17910, Andrew385, Kikugawa134520}, which indicates that the   developed fit properly describes the dispersion of the bulk bands in the    vicinity of the FL.  The  integral of the DOS up to the FL   gives the electron count of  3.9  electrons per BZ, which again, within the   experimental accuracy, agrees with the expected 4 electrons per BZ.

Electronic structure of materials is frequently discussed in terms of effective
masses. Such a reduction to a single integral value $m^*$ enables a comparison
between various experimental and theoretical methods. In this context, it is useful
to remind that besides the band mass tensor $m_{\mu,\nu} =
\hbar^2(\frac{\partial^2\epsilon(\mathbf{k})}{\partial k_\mu \partial k_\mu})^{-1}$
different effective masses are frequently considered: (1) the band mass
$m_\mathrm{b}$ as it can be obtained from the bare electronic dispersion, (2) the
thermodynamic mass $m^*$, (3) the cyclotron resonance mass $m_\mathrm{c}$,  (4) the
susceptibility mass $m^*_\mathrm{suscept}$,  (5) the plasma frequency mass
$m_\mathrm{p}$\,\cite{Bergemann639, Merino2416}. Here, the quasiparticle densities of states can be easily recalculated into thermodynamic masses ($m^*/m_\mathrm{e} = \pi\hbar^2 g/ m_\mathrm{e} a^2 $) and compared, on the band by band
basis, to the thermodynamic masses reported in de Haas--van Alphen measurements as
summarized in the table below.
\begin{table}[h]
\begin{tabular}{l c c c c c }
\firsthline mass type &  $\alpha$ &
$\beta$ & $\gamma$ & total &year \\
\hline
(this study)                               & 5.4 & 4.8& 16.7 & 26.9\\
\hline
Cyclotron thermodynamic\,\cite{Bergemann2001371, Bergemann639}    & 3.3 & 7.0
& 16.0 & 26.3&2001\\
Cyclotron thermodynamic\,\cite{Andrew385}       & 3.4 & 7.5 & 14.6 & 25.5&1998\\
Cyclotron thermodynamic\,\cite{Mackenzie3786}   & 3.4 & 6.6 & 12. & 22.0&1998\\
Cyclotron thermodynamic\,\cite{Mackenzie1996510}& 3.2 & 6.6 & 12.0 & 21.8&1996\\
\hline
Cyclotron resonance\,\cite{Bergemann639} & 2.1 & 4.3 & 5.8 & 12.2& 2003\\
Cyclotron resonance\,\cite{Hill3374} & 4.3 & 5.8 & 9.7 & 19.8 & 2000\\
\lasthline
\end{tabular}
\end{table}

As one may notice there is a gradual increase of the total thermodynamic mass
reported by the de Haas--van Alphen measurements with time, which is probably
related to the improving quality of the available crystals. Our total mass
(26.9\,$m_\mathrm{e}$) is closest to the most recent dHvA value of
26.3\,$m_\mathrm{e}$. Similarly, we find a good correspondence for the mass of the
$\gamma$ band, however there seems to be a difference in the masses of $\alpha$ and
$\beta$ bands. While in the current fit these bands have practically the same mass,
in the de Haas--van Alphen data their mass ratio is about two. In the
two-dimensional case the density of states, and hence the effective mass, is
inversely proportional to the Fermi velocity and directly proportional to the length of the Fermi surface contour. Therefore, assuming equal velocities for the $\alpha$ and $\beta$ band would give a mass of the $\beta$ band to be twice that of the $\alpha$ band, as the average radius of the $\beta$ band is about twice as high. As clarified in ref. \onlinecite{Bergemann639} (p.\,686) this assumption was `actually used ... as a guiding line' when extracting the susceptibility effective mass. Nevertheless, as can be seen from Fig.\,3a, the quasiparticle Fermi velocity of the $\alpha$ band is systematically lower than that for the $\beta$ band, thus the accurate account of the variation of the Fermi velocity, $v_\mathrm{F}$, and $k_\mathrm{F}$ yields about the same thermodynamic masses for the $\alpha$ and $\beta$ bands. We expect that the value extracted from the presented TB model correctly reproduces the relation between the effective mass of the $\alpha$ and $\beta$ bands.

Besides the heat capacity, the obtained effective TB model can be used
to calculate other averaged properties over the Brillouin zone, such as the plasma
frequency, whose value is given by
\begin{equation}
\hbar \Omega_{xx} =
\sqrt{\sum\limits_{i=\alpha, \beta, \gamma}\frac{e^2}{L_a L_b L_c \epsilon_0}
\left\langle f_k \frac{\partial^2 E^{(i)}}{\partial k_x \partial k_x}
\right\rangle_\textup{\tiny{\!BZ}} },
\end{equation}
from where we get $\hbar \Omega_{xx}$ = 1.3\,eV. As expected the value is about 4 times smaller than the one obtained based on the unrenormalized band structure calculation\,\cite{Singh1358}.

In conclusion, we have developed an effective tight-binding model that is able to
capture the low energy electronic features of the quasiparticle spectra of
{\sample} taking ARPES data as an input. Owing to different degree of
renormalization in the bulk and at the surface, the bulk bands have been properly
selected and analyzed for the determination of the quasiparticle model. We have
extracted the momentum and orbital dependance of the Fermi velocity and of the
effective masses close to the FL. As a demonstration of the use of the
derived model, we have calculated the density of states and found a good agreement
between the Sommerfeld coefficient calculated based on the obtained fit and the one
directly measured. We believe that the developed model can be of value for a more
realistic input to compute the orbital dependent magnetic properties in order to
test, for example, the relevance of the ferromagnetic or antiferormagnetic
fluctuations in settling the spin-triplet pairing in the superconducting phase of
{\sample}\,\cite{Taylor137003, Raghu136401, Puetter27010}.

The work was supported  by DFG grant ZA 654/1-1.   E.\,C. and B.\,P.\,D. thank the
Faculty of Science at the University of Johannesburg for travel funding. M.\,C.,
R.\,F. and A.\,V. acknowledge support and funding from the FP7/2007-2013 under
grant agreement N.264098-MAMA.

\end{document}